\documentclass[preprint,12pt,authoryear]{elsarticle}

\usepackage[T1]{fontenc}
\usepackage[utf8]{inputenc}
\usepackage{color}
\usepackage{babel}
\usepackage{mathtools}
\usepackage{array}
\usepackage{amsmath}
\usepackage{booktabs}
\usepackage{pdflscape}
\usepackage{amssymb}
\usepackage{rotating}
\usepackage{graphicx}
\usepackage{epstopdf}
\usepackage{hyperref}
\hypersetup{
	colorlinks=true,
	linkcolor=blue,
	citecolor=blue,
	urlcolor=blue
}
\graphicspath{{./Figures/}}

\allowdisplaybreaks

\journal{Chaos Solitons and Fractals}

\begin{document}

\begin{frontmatter}



\title{Rare Events and Redundancy in Random Walkers Target Search in a Finite Domain}

\author[label1]{Elisabetta Ellettari} 

\author[label1]{Giacomo Nasuti}

\author[label1,label2]{Alberto Bassanoni}

\author[label1,label3]{Alessandro Vezzani}

\author[label1,label2]{Raffaella Burioni}

\cortext[cor1]{Corresponding author Alessandro Vezzani: alessandro.vezzani@unipr.it}

\affiliation[label1]{organization={Dipartimento di Scienze Matematiche, Fisiche e Informatiche,
Universit\`a degli Studi di Parma},
            addressline={Parco Area delle Scienze, 7/A}, 
            city={Parma},
            postcode={43124}, 
            country={Italy}}

\affiliation[label2]{organization={INFN, Gruppo Collegato di Parma},
            addressline={Parco Area delle Scienze, 7/A}, 
            city={Parma},
            postcode={43124}, 
            country={Italy}}

\affiliation[label3]{organization={Istituto dei Materiali per l'Elettronica ed il Magnetismo (IMEM-CNR)},
            addressline={Parco Area delle Scienze, 37/A}, 
            city={Parma},
            postcode={43124}, 
            country={Italy}}

\begin{abstract}
Finding a target in a complex environment is a fundamental challenge across natural systems, from chemical reactions to sperm cells reaching an egg. A powerful strategy to reduce search times is redundancy: deploying many independent searchers increases the probability that at least one succeeds, particularly when success is driven by rare events. When the underlying stochastic motion features broadly distributed step lengths, rare long relocations dominate the dynamics, making redundancy especially effective. Here, we investigate the statistics of extreme events for the mean first passage time in a system of $N$ independent walkers performing power-law–distributed jumps with finite velocity, where target-reaching events are governed by single large fluctuations. We show that the mean first passage time of the fastest walker scales as $\langle T_N \rangle \sim 1/N$, representing a dramatic speed-up compared to classical Brownian motion, and saturates at the minimum value $X/v$. We further extend the model to include random velocity. For fixed $N$, we identify a crossover, governed by a critical tail exponent $\alpha_c$, separating a regime dominated by a single large fluctuation (“big jump”) from a regime characterised by Gaussian extreme-value statistics arising from finite sampling effects. From these results, we derive a scaling law that links the number of walkers $N$ to the size $X$ of the search region. Our results demonstrate how redundancy, combined with rare-event statistics, can efficiently organise target-search processes in complex biological environments. As a prototypical example, we consider mammalian fertilisation and derive, within a coarse-grained description, a cross-species scaling relation between the number of spermatozoa and the typical uterine size.
\end{abstract}


\end{frontmatter}

\section{Introduction}


The problem of target reaching plays a central role in stochastic processes, from chemical reactions to ecological spread, intracellular signaling, and fertilization. The mean first passage time (MFPT) \cite{First_Passage_1, First_Passage_2, First_Passage_3} sets the fundamental timescale to trigger the reaching events, making its reduction a key goal in both natural and engineered systems. A natural strategy to accelerate target finding is to use redundancy, that is, deploy many searchers in parallel. Understanding how redundancy shapes the statistics of rare threshold-crossing events by $N$ agents becomes crucial in many-body and collective search problems. However, this approach incurs a cost, and identifying the optimal number of searchers becomes a fundamental question in the design of efficient search strategies.

Many theoretical studies have focused on Brownian walkers, where the extreme MFPT, that is, the average time for the fastest among $N$ agents to hit first the target, is described by narrow escape theory \cite{N_Brownian_1, N_Brownian_2, N_Brownian_3_ReviewHolcman, N_Brownian_4}. In such cases, its scaling $\langle T_N \rangle \sim 1/\log N$ with the number of walkers $N$ \cite{N_Brownian_Lawley_1, N_Brownian_Lawley_2, N_Brownian_Lawley_4} reveals a slow logarithmic decrease, requiring a large number of searchers to achieve modest gains in search time and indicating that redundancy is only weakly exploited in Brownian search processes.
Non-Gaussian dynamics improve performance, even for a single particle. For example, intermittent search processes and two-state models feature an alternation between slow, local diffusion, and fast, ballistic relocation phases \cite{TwoStateModels_1, TwoStateModels_2, TwoStateModels_3, TwoStateModels_4, TwoStateModels_5_Levy_vs_2State, TwoStateModels_6_Levy_vs_2State}. These models are particularly effective in heterogeneous environments, yet still exhibit a logarithmic extreme MFPT scaling with $N$ in unbounded domains and, in confined settings, their behavior becomes strongly geometry-dependent \cite{N_Brownian_Lawley_1, N_Brownian_3_ReviewHolcman}. Another strategy is to exploit fluctuating diffusivities. In diffusing-diffusivity models, where the diffusion coefficient varies stochastically, the extreme MFPT scales faster, as $\langle T_N \rangle \sim 1/(\log N)^\beta$ with $\beta > 1$ \cite{N_Brownian_5_Sposini, N_Brownian_6_Sposini}. This enhanced scaling has been linked to biological scenarios such as sperm motility and polymer diffusion in crowded media \cite{N_Brownian_NonGauss_1}. Finally, another way to enhance the extreme MFPT is to exploit heterogeneity in the initial conditions of the searchers. For instance, randomness in the initial spatial distribution of independent walkers can significantly affect the statistics of the fastest arrival events in diffusion-controlled reactions \cite{N_walkers_reaction_diffusion_1} .

In this work, we investigate the statistics of the extreme mean first-passage time (MFPT) in a broad class of systems composed of $N$ independent heavy-tailed (HT) walkers. Our focus is on how redundancy controls extreme first-passage events in finite domains. The walkers move at constant speed and perform jumps whose lengths $x$ are drawn from a power-law distribution $x^{-(1+\alpha)}$. For $\alpha < 2$, the resulting dynamics is superdiffusive and corresponds to L\'evy walks, where displacements occur at finite speed while the jump length distribution has infinite variance 
\cite{LevyWalks_1, LevyWalks_2, PRE_Eli_Levy, PRL_Eli_Levy}. In contrast,  for $\alpha > 2$ the dynamics crosses over to standard diffusion. This framework therefore captures both anomalous and normal diffusion regimes commonly observed in natural systems. Such processes are central to search and foraging strategies and have been reported across a wide range of systems, from bacteria \cite{LevyWalksAppl_Bact1} and animals \cite{LevyWalksAppl_Anim1} to human mobility patterns \cite{LevyWalksAppl_Human1}.

In these heavy-tailed systems, rare events, and consequently extreme MFPTs, are typically dominated by a single large fluctuation, in accordance with the big jump principle (BJP) \cite{BJ_theorem, BJ_Foss, BJ_1, BJ_2, BJ_3, BJ_4, BJ_5, BJ_6, BJ_Omer, BJ_subexp, BJ_subexp_2, BJ_subexp_3}. By explicitly estimating the contribution of these dominant fluctuations, we derive the extreme MFPT for a system of $N$
walkers and show that it scales as $\langle T_N \rangle \sim 1/N$, saturating at the physical lower bound $X/v$. This inverse scaling reflects a substantial speed-up due to redundancy, markedly enhancing search efficiency compared to classical diffusive scenarios and other search models. 

We exploit this result to determine the optimal level of redundancy, quantified by the number of HT walkers required to reach a target within a prescribed MFPT in a finite one-dimensional domain of size $X$. Remarkably, starting from a rare-event analysis of a single walker, our framework enables us to reconstruct the full probability density function (PDF) of the first-passage time for a system of 
$N$ walkers in the appropriate asymptotic limit. All calculations are carried out for walkers moving at constant velocity; however, the approach can be straightforwardly extended to random velocity models, paving the way for natural generalisations to higher dimensions.

As a concrete application, we consider the biological problem of sperm reaching the egg as a paradigmatic example of a redundancy-controlled target search process. Although sperm motion appears diffusive on short timescales \cite{SpermMotility_1, SpermMotility_2, SpermMotility_3, SpermMotility_4}, experimental evidence of directional persistence and motility heterogeneity \cite{SpermMotility_RandomAngleBallistic, SpermMotility_TwoStateModel_1, SpermMotility_TwoStateModel_2} suggests the presence of heterogeneous transport mechanisms, which can be effectively captured at a coarse-grained level by HT processes. In particular, we are able to capture a scaling relation between the number of spermatozoae required for fertilisation and the size of the uterus in different mammalian species. These findings show that a redundancy-based extreme-statistics framework is compatible with the empirically observed cross-species scaling relations, without requiring explicit consideration of the microscopic dynamics governing sperm motility. 
Beyond biological search, first-detection and extreme-arrival statistics also emerge in transport problems driven by strong fluctuations, such as tracer detection in turbulent flows \cite{N_walkers_turbulence}.


\section{Rare hitting times for $N$ Heavy-Tailed Walkers}

We consider a system of $N$ independent one-dimensional HT walkers, representing a redundant population of searchers. Each walker undergoes a continuous-time jump process at constant speed $v$, with jump durations drawn from a power-law distribution:
\begin{equation}
\label{eq:single_jump}
p(t) =
\begin{cases}
\frac{\alpha t_0^{\alpha}}{t^{1+\alpha}} & t>t_0 \\
0 & t<t_0.
\end{cases}
\end{equation}
with $t_0$ being a time cutoff and speed directions chosen randomly. Our goal is to compute the probability $\text{Prob}(T,X,N)$ that at least one out of the $N$ walkers reaches a fixed target at distance $X$ within a time shorter than $T$. We denote by $\ell(T)\sim T^{1/z}$ the characteristic scaling length of the process. This means that for times $T\sim X^z$, the probability of reaching a target at distance $X$ within time $T$ is expected to depend on the scaling variable $T/X^z$. For $\alpha>2$, one has $z=2$, corresponding to diffusive motion governed by the Central Limit Theorem. In contrast, for $1<\alpha<2$, the dynamics is superdiffusive with dynamical exponent $z=\alpha$ \cite{LevyWalks_1}.

{The BJ estimate provides a way to evaluate the probability of rare events at large distances $X\gg \ell(T)$ or equivalently at short times $T\ll X^z$ in systems where the distributions of the step lengths or durations is sub-exponential. Originally developed as a heuristic tool for estimating probability density functions, the BJ principle has more recently been extended to first-passage problems as hitting times \cite{BJ_5} and record statistics \cite{BJ_6}.
Interestingly, in several contexts (\cite{BJ_Foss, PRE_Eli_Levy, PRL_Eli_Levy, BJ_Zamparo, BJ_LenciBianchi}) this heuristic approach has been shown to yield asymptotically exact results. The BJ principle states that that rare events, such as fast exit times with $T\ll X^z$, are typically realized through a single long and fast jump that dominates the dynamics.  In particular, for a single walker, the hitting probability $\text{Prob}(T,X)$ of a target at distance $X$ is determined,  in the regime $T \ll X^d$, by a single big jump of duration longer than $X/v$, while all other jumps contribute negligibly to the total displacement \cite{BJ_1,BJ_3,BJ_5,BJ_6}. Therefore, $\text{Prob}(T,X)$ can be estimated as:
\begin{align}
\label{probwalk}
\text{Prob}\left(T,X \right)&\sim \frac{1}{2} \langle n_T\rangle \times \text{Prob}\left(t > X/v\right) \\ \nonumber
&\sim 
\begin{cases}
  \frac{T-X/v}{2 \langle t \rangle} \int_{X/v}^{\infty} dt \ p(t)  & \text{if $T> X/v$}\\
  0 & \text{if $T\leq X/v$}
\end{cases}
\end{align}
where $\text{Prob}(t > X/v)$ is the probability of drawing a step larger than $X$, $\langle n_T \rangle$ is the average number of attempts to make the big jump and the factor $1/2$ accounts for the probability of escape from one side only, assuming $X$ positive (this factor is absent in the case of a two-sided escape). Then $\langle n_T \rangle=(T-X/v)/\langle t \rangle$, where $\langle t \rangle = \int dt' \ t' p(t') =\frac{\alpha}{\alpha-1}t_0$ is the average duration of a single step and $T-X/v$ is the available time to perform the big jump. 

Since all $N$ walkers are independent, the probability that at least one of them can reach $X$ during a total time $T$ is the complementary probability. Therefore, in the big jump regime $T\ll X^z$:

\begin{figure}
\begin{center}
\includegraphics[scale = 0.28]{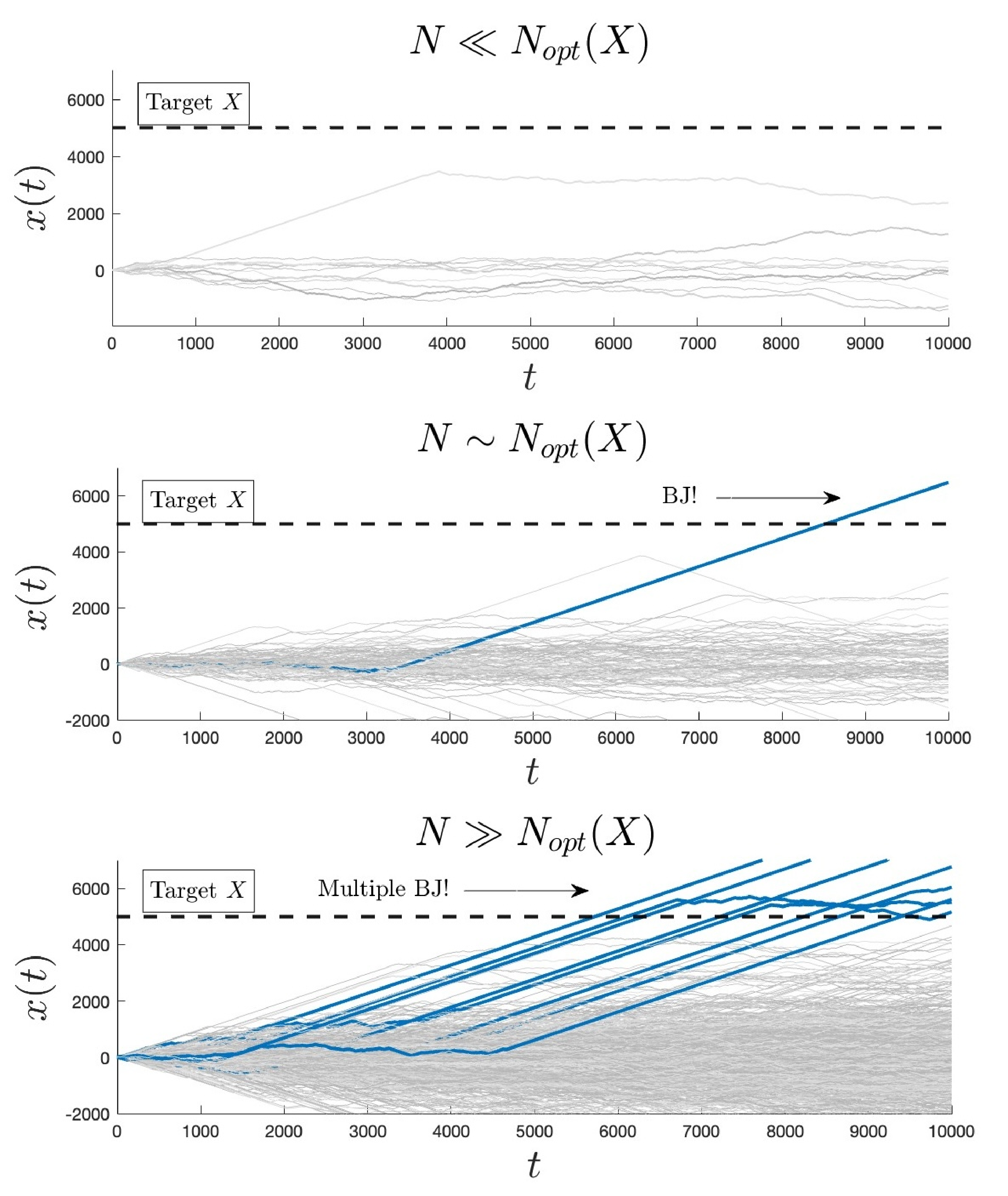}
\end{center}
\caption{Sample trajectories of 
$N$ independent one-dimensional HT walks with target set at 
$X=5000$ (dashed black line), shown for the three possible regimes of walker redundancy. In the plot on top, i.e. the low redundancy regime, target reaching is a rare event due to insufficient sampling. In the middle plot, i.e. the optimal redundancy regime, typically the target is reached by a single walker performing a big jump. In the bottom plot, i.e. the high redundancy (or saturated) regime, many walkers reach the target, so that the fastest performs the jump at $t=0$ and the exit times is $T=X/v$. Parameters are $t_0=1$, $v=1$ and $\alpha=1.3$;}
\label{fig:levy_regimes}

\end{figure}

\begin{align}
\label{probN}
    \text{Prob}(T,X,N) &\sim 1 - (1 - \text{Prob}(X,T))^N \\ \nonumber
    &\sim 1 - \exp\left[-\frac{N}{N_{opt}(X)}f\left(\frac{vT}{X}\right)\right], 
\end{align}
for  $X<vT$ and $\text{Prob}(T,X,N)=0$ for  $X>vT$. In \eqref{probN} $f(vT/X) = (\frac{vT}{X}-1)$ is a scaling function that captures the asymptotic behaviour in the rare event regime, dominated by ballistic motion. Within this picture, redundancy plays a purely statistical role: increasing the number of walkers increases the probability that at least one trajectory realizes the rare fluctuation required to reach the target. So, the quantity $N_{opt}(X)$ defines an optimal number of walkers required to efficiently sample rare big-jump events at a distance $X$:
\begin{equation}
\label{N_opt}
N_{opt}(X) =\frac{2\alpha}{\alpha-1}\left(\frac{X}{vt_0}\right)^{\alpha-1}.
\end{equation}

This quantity defines three different regimes, which are resumed in Figure \ref{fig:levy_regimes}:

\noindent
\textbullet\ Low-redundancy regime ($N \ll N_{opt}(X)$): The number of particles is insufficient to compensate for the rarity of large jumps, leading to poor sampling of extreme events. In this case, the total probability of reaching the target scales linearly with $N$, and we recover the single-walker result multiplied by $N$, i.e. $\text{Prob}(X,T,N) \sim N \cdot \text{Prob}(X,T)$;

\noindent
\textbullet\ Optimal-redundancy regime ($N \sim  N_{opt}(X)$): The number of particles is just sufficient to ensure that rare big-jump events are effectively sampled, so that target reaching is typically dominated by a single walker performing a large fluctuation. In this scaling regime, our theoretical approach accurately captures the total probability for $N$ walkers, as all successful events are governed by the BJP. Indeed, starting from a rare events calculation of a single particle, we reconstruct the full PDF of the system of $N$ particles. This allows us to compute key observables, such as the extreme MFPT, using Eq.~\eqref{probN};

\noindent
\textbullet\ High-redundancy regime ($N \gg N_{opt}(X)$): The number of walkers is more than sufficient, and the system becomes saturated. With high probability, at least one walker reaches the target almost instantly. The probability approaches a step function:
 i.e. $\text{Prob}(X,T,N) \to 1 \ \text{for } X < vT$ and $\text{Prob}(X,T,N) =0 \ \text{for } X > vT$;

Notice that for $\alpha\le 1$ the scaling length of the process becomes ballistic, $\ell(T)\sim T$, and comparable with the long jump itself. In this case, the BJP cannot be applied and a different approach is needed \cite{BJ_1}.

\section{The Extreme First Passage Time}

From Eq. \eqref{probN}, taking a derivative with respect to $T$ yields the exit time probability distribution:
\begin{align}
\label{PDF_N_T}
P_N(T) &= \frac{d}{dT}\text{Prob}(X,T,N) \\ \nonumber
&= \frac{N}{N_{opt}(X)} \left( \frac{v}{X} \right) \exp \left[ - \frac{N}{N_{opt}(X)} f\left(\frac{vT}{X}\right) \right] 
\end{align}
for $X\leq vT$, while $P_N(T\leq X/v)=0$ as a consequence of the finite propagation speed of the walkers.

In both the optimal and high-redundancy regimes, $\text{Prob}(T,X,N)$  is of order one for $X\sim vT$. This indicates that, in these regimes, the same rare-event mechanism that governs the tail of the distribution also controls typical exit events. Although the expression for $P_N(T)$ is derived from a tail estimate, it effectively captures the full exit-time statistics, since the dynamics is dominated by single big-jump events selected through redundancy. Consequently, this expression can be reliably used to compute the MFPT. Moreover, we observe that the probability density function decays exponentially with $T$, reminiscent of the Brownian case. In contrast to Brownian motion however, the decay rate is here governed by the big jump statistics, directly reflecting the heavy-tailed nature of the process.

The follows from the first moment of $P_N(T)$:
\begin{align}
\langle T_N \rangle &= \int_{X/v}^\infty T \cdot P_N(T) \, dT = \frac{X}{v} + \frac{2 \langle t \rangle}{N} \left( \frac{X}{v t_0} \right)^\alpha.
\label{eq:MFPT}
\end{align}
This result is the sum of a ballistic contribution $X/v$, reflecting the minimal time needed to perform a single big jump of size $X$, and a statistical correction proportional to $1/N$. Compared to the classical $ \langle T_N \rangle \sim 1/\log N $ scaling of Brownian walkers (for a complete derivation see \ref{AppendixA}), Eq.(\ref{eq:MFPT}) reveals a dramatic enhancement of search efficiency. 

\begin{figure}
\begin{center}
\includegraphics[scale = 0.60]{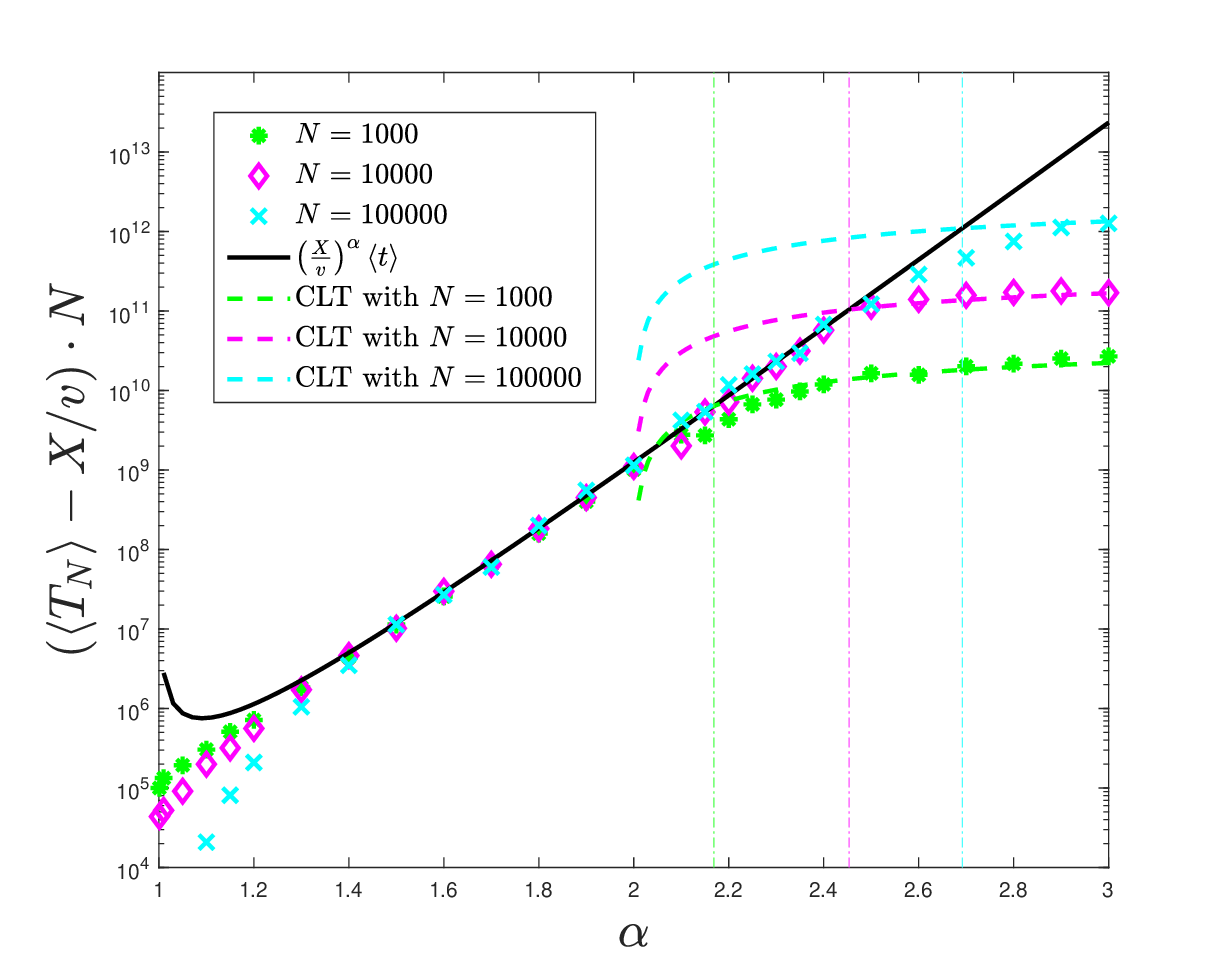}   
\end{center}
\caption{The rescaled extreme MFPT $\langle T_N \rangle$ plotted as a function of the power law exponent $\alpha$ at different values of $N$ in
logarithmic scale. The black straight line corresponds to our
theoretical prediction for HT walks \eqref{eq:MFPT}. The other colored
dotted lines correspond to the logarithmic behaviour given by the CLT of Brownian motion for $\alpha>2$.
We can see that, increasing $N$, the crossover point to the CLT at $\alpha_c(N)$ moves at larger $\alpha$, here represented by the vertical colored dotted axes. Here $t_0=1$ , $v = 1$ and $X = 25000 v t_0$, which is the
typical scale of human uterus longitudinal size. As a final remark, for a sufficiently high number of searchers $N$ in the case of $\alpha>2$ one can notice that the extreme MFPT in the Brownian dynamics is typically order of magnitude higher that in our HT dynamics, see for example $\alpha = 2.2$ and $N=100000$. All the numerical simulations are performed in a finite domain with two absorbing boundaries at $X$ and $-X$, and the extreme FPT is measured directly as the earliest arrival time among the $N$ walkers.}
\label{fig:tau_N_vs_alpha}
\end{figure}

Figure \ref{fig:tau_N_vs_alpha} represents our core result, as it summarises how redundancy, extreme statistics, and finite-size sampling limitations jointly shape the extreme MFPT of many HT walkers. It shows $\langle T_N\rangle$ as a function of the tail exponent $\alpha$, for different fixed values of the number of walkers $N$. 

At fixed $N$, $\langle T_N \rangle$ exhibits a clear crossover between two qualitatively distinct behaviors. 
For sufficiently small $\alpha$, the extreme MFPT follows the ballistic scaling predicted by the BJP, i.e. $\langle T_N\rangle \sim 1/N$, showing that target reaching is dominated by a single rare, long relocation event. 
For larger $\alpha$, the numerical data progressively deviate from this prediction and approach the logarithmic scaling $\langle T_N\rangle \sim 1/\log N$ characteristic of Brownian motion \cite{N_Brownian_Lawley_1, N_Brownian_Lawley_2, N_Brownian_Lawley_4}.
This crossover is controlled by a critical value $\alpha_c(N)$, which explicitly depends on the number of walkers. It is obtained by matching the asymptotic expression of the extreme MFPT in the big-jump regime, Eq. \ref{eq:MFPT}, with the corresponding Brownian regime (see \ref{AppendixA}). Solving the resulting transcendental equation yields:
\begin{equation}
 \alpha_c(N) = \frac{W_0(\mathcal{C}_{X,v,t_0}(N,\alpha_c))}{\ln \left(\frac{X}{vt_0}\right)}
\label{alpha_c_N}
\end{equation}
where $W_0(.)$ stands for the principal branch of the Lambert function. Its argument $\mathcal{C}_{X,v,t_0}(N,\alpha_c)$ depends on $N, \alpha_c$ and the other parameters of the model, and is given by:
\begin{equation}
\mathcal{C}_{X,v,t_0}(N,\alpha_c)=\frac{N}{2}\left(\frac{X}{vt_0}\right)\ln \left(\frac{X}{vt_0}\right)\left[\frac{(\alpha_c-2)}{2\ln\left(\frac{N}{\sqrt{\pi}}\right)}\left(\frac{X}{vt_0}\right) - 1\right]
\end{equation}

By construction, $\alpha_c(N)>2$ for finite $N$, consistent with the fact that Gaussian behaviour can only emerge when the variance of the step-length distribution is finite. In the large $N$ limit one can show that the crossover value increases logarithmically, i.e. $\alpha_c(N) \sim \ln N$. The full derivation of $\alpha_c(N)$ and of its logarithmic behaviour can be found in \ref{AppendixB}.  
Physically, this behavior reflects a sampling limitation: for sufficiently large $\alpha$, the tails of the distribution become too thin to be efficiently sampled even by $N$ particles. In this regime, tipically target reaching is no longer controlled by a single dominant fluctuation, but by the accumulation of many moderate steps, leading to Gaussian extreme statistics and logarithmic scaling of $\langle T_N\rangle$.
This interpretation naturally connects with the concept of an optimal number of walkers discussed in the previous section. 

\begin{figure}
\begin{center}
\includegraphics[scale = 0.55]{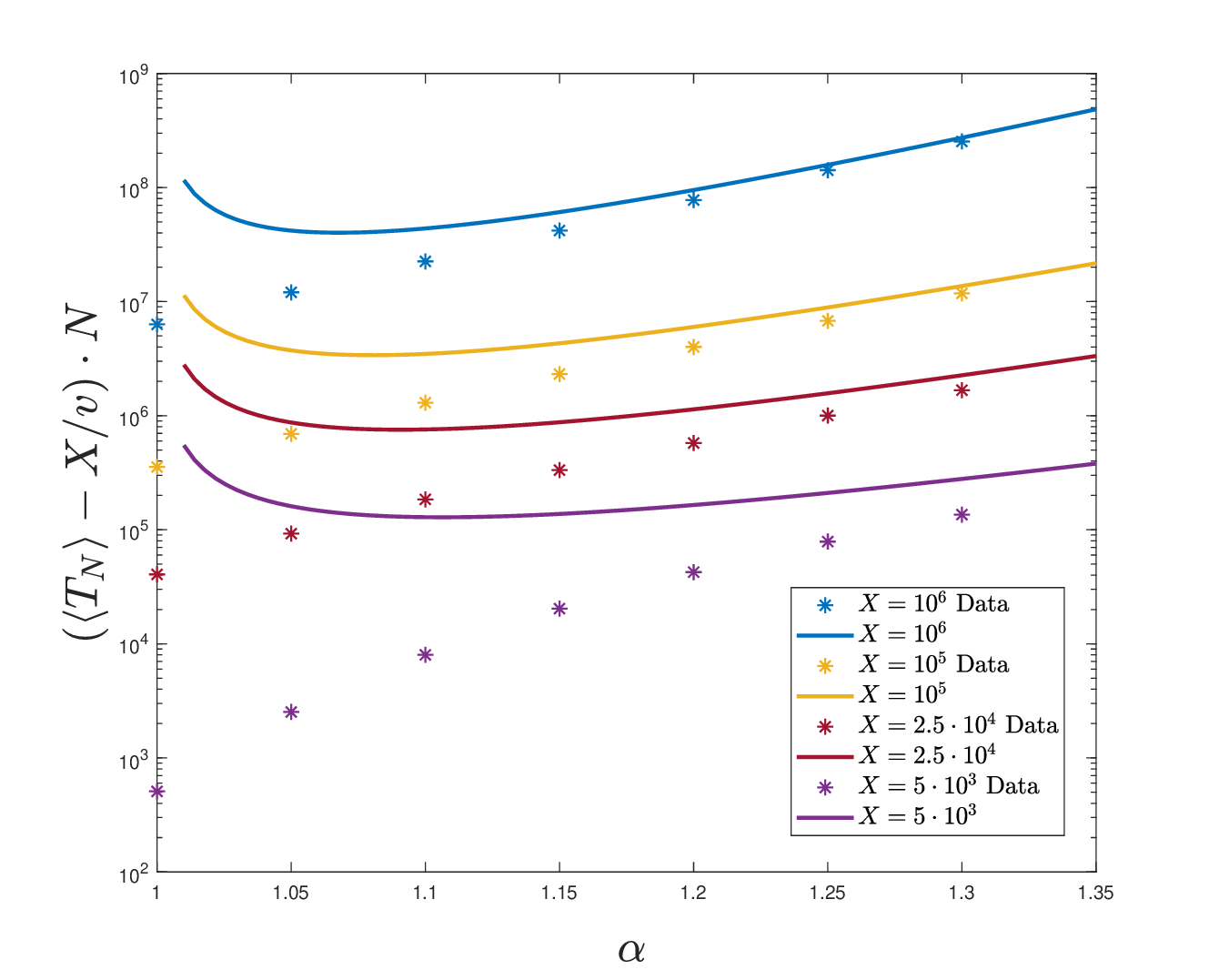}
\end{center}
\caption{Logarithmic plot of normalized the extreme MFPT $(\langle T_N \rangle - X/v) \cdot N$ as a function of the target position $X$ when the values of $\alpha \rightarrow 1^+$. The colored stars are data calculated by simulating numerically the extreme MFPT of $N=10000$ for different values of $\alpha$ and $X$. The colored straight lines are the theoretical prediction for the extreme MFPT, which tends to diverge when $\alpha \rightarrow 1^+$. The dotted colored vertical axes corresponds to the asymptotical value of $\alpha(X)$ under which the BJP starts to fail. In these simulations, the product of velocity and of the typical waiting time is fixed to $vt_0 = 1$ and the asymptotic constant is fixed to $C=2.2$;}
\label{fig:alpha_near_1}
\end{figure}

A second, distinct finite-size effect appears in Figure \ref{fig:tau_N_vs_alpha} in the opposite limit $\alpha\to1^{+}$, where deviations between numerical simulations and theoretical predictions are also observed within the big-jump regime. 
This effect has a different physical origin and is not related to the crossover to Gaussian behavior. When $\alpha\to1^{+}$, the typical length scale explored by the walk over a time $T$, $\ell(T) \sim T^{1/\alpha}$,
becomes comparable to the ballistic length scale associated with a single big jump. 
In particular, at the expected exit time $T\sim X/v$, one has $\ell(T)\sim X$. 
Since rare long jumps are no longer significantly larger than typical fluctuations, the basic assumption underlying the BJP — a separation length scale between typical and rare events — ceases to hold.

The validity of the BJP as $\alpha\rightarrow 1^+$ can be recovered by increasing the target distance $X$, restoring a separation of length scales. A quantitative criterion for this regime is provided by this lower bound for the exponent $\alpha$:
\begin{equation}
\label{alpha_near_1}
        \alpha(X) \underset{\alpha\rightarrow 1^+}{>} \frac{\ln \left(\frac{X}{vt_0}\right)}{\ln \left(\frac{X}{vt_0}\right) - C},
\end{equation}
where $C\sim O(1)$ is an arbitrary constant which depends by the parameters of the power tailed distribution. Its full derivation can be found in \ref{AppendixC}.
This explains why deviations disappear when the target is moved sufficiently far from the origin. This can be observed more clearly in Figure \ref{fig:alpha_near_1}.

In summary, two distinct finite-size effects are present: 
an $\alpha_c(N)$ crossover driven by sampling limitations at fixed $N$, leading from HT to Brownian extreme statistics; an $\alpha(X)$ breakdown of the BJP caused by the loss of scale separation between typical fluctuations and ballistic jumps. 
Both effects delimit the range of validity of the ballistic $1/N$ scaling of the extreme MFPT $\langle T_N \rangle$ for $N$ HT walkers.

\begin{figure}[t]
\centering
\includegraphics[width=0.8\textwidth]{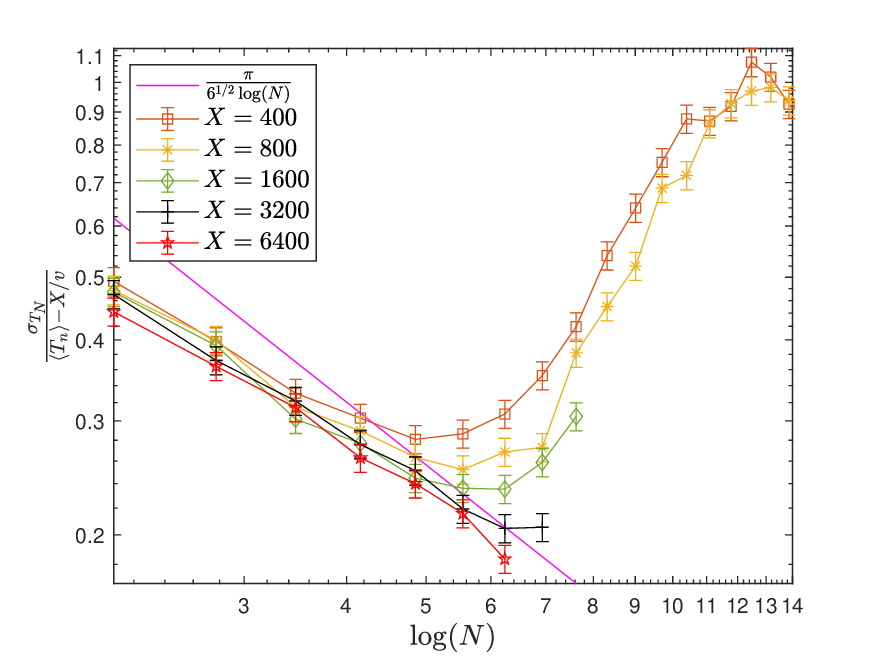}
\caption{
Ratio between the standard deviation of the extreme first-passage time and the extreme MFPT minus the ballistic lower bound, $\frac{\sigma_{T_N}}{\langle T_N \rangle - X/v}$, plotted as a function of the redundancy level $N$ for $\alpha=2.8$ and different target distances $X$. The magenta straight line corresponds to the Brownian prediction $\sim (\ln N)^{-1}$. At low redundancy, the fluctuations decrease slowly with $N$, consistently with diffusive extreme-value statistics. As $N$ increases toward the optimal redundancy regime, the ratio approaches a constant value of order one, in agreement with the big-jump prediction \eqref{fuct_opt}. Increasing the system size $X$ extends the regime where the Brownian scaling is observed before the crossover to the big-jump dominated regime.
}
\label{fig:fluctuations}
\end{figure}

As a final remark, the full distribution allows us to characterize not only the mean exit time but also its fluctuations. In particular, Eq. \eqref{PDF_N_T} directly yields the standard deviation of the exit time for $N$ walker: $\sigma_{T_N}= \frac{2 \langle t \rangle}{N} \left( \frac{X}{v t_0} \right)^\alpha$. We estimate the size of the fluctuation  as the ratio between the standard deviation and the difference between the average and the minimum exit times (i.e. $\langle T_N \rangle-X/v$). In the  optimal regime this gives:
\begin{equation}
\label{fuct_opt}
        \frac{\sigma_{T_N}}{\langle T_N \rangle-X/v}=1.
\end{equation}
On the other hand, for $N\ll N_{opt}(X)$ and $\alpha>2$, the exit time is governed by diffusive (Brownian) motion. In this regime, the standard deviation of the exit time at large $N$ behaves as \cite{N_Brownian_Lawley_2}: $\sigma_{T_N} \approx \frac{\pi}{6^{1/2} (\ln N)} \langle T_N \rangle$. Since in the low abundance regime $\langle T_N \rangle\sim X^z\gg X/v$, this implies:
\begin{equation}
\label{fuct_brownian}
        \frac{\sigma_{T_N}}{\langle T_N \rangle-X/v}\approx \frac{\pi}{6^{1/2} (\ln N)}.
\end{equation}
Eq. \eqref{fuct_brownian} shows that, in the low redundancy regime, fluctuations decrease as the number of walkers increases. In contrast, in the optimal regime, where rare events control the exit time, fluctuations remain comparable to the mean independently of the number of walkers as indicated by Eq. \eqref{fuct_opt}.

The crossover between these two fluctuation regimes is illustrated in Figure \ref{fig:fluctuations}, where the ratio between the standard deviation and the excess MFPT is plotted as a function of the redundancy level $N$. At low redundancy, the fluctuations decrease slowly with $N$, consistently with the Brownian logarithmic prediction. By contrast, in the optimal redundancy regime the ratio approaches a constant of order one, reflecting the persistence of broad fluctuations generated by the big-jump mechanism.

\section{Extension to Random Velocity Models}

In the theoretical framework introduced above, walkers were assumed to move with fixed velocity $v$. 
This assumption allows for a transparent analytical treatment and highlights the role of rare, long relocation events in determining the extreme first-passage statistics. 
Importantly, the validity of the BJ does not hinge on this assumption but only on the sub-exponential distribution of step lengths or durations. 
However, the specific form of the probability of rare events depends on the microscopic dynamics of the single step. For example accelerated processes, can also be analyzed within generalized big-jump frameworks \cite{BJ_3}. In particular, the wait-then-jump and jump-than-wait processes can be interpreted as singular limits of such accelerated step  dynamics.

In many physical and biological systems, individual agents exhibit a distribution of effective velocities due to intrinsic heterogeneity, orientation dispersion, and interactions with a complex environment. We therefore consider a population of $N$ independent walkers that perform steps with durations drawn from the same power-law waiting-time distribution as in the constant-velocity case, but now move with velocities $v$ independently sampled from a prescribed probability density $f(v)$. The velocity is taken to be constant within each step, while differing across walkers.
For simplicity, and in accordance with directed target-reaching scenarios, we focus on one-sided velocity distributions with support on $v>0$.

Within the big-jump regime, target reaching at distance $X$ within a time $T$ is dominated by a single long step, whose duration is of order $X/v$. 
Conditioned on a given velocity $v$, the probability that a single walker reaches the target within time $T$ can therefore be estimated using the same approach as in the constant-velocity case. 
Averaging the velocity distribution and exploiting the independence of walkers, the probability that at least one out of $N$ walkers reaches the target reads
\begin{equation}
\mathrm{Prob}(X,T,N)
\simeq
1-\left[
\int_0^{\infty}
\bigl(1-\mathrm{Prob}(X,T\mid v)\bigr)
\,f(v)\,dv
\right]^N .
\end{equation}

In the rare-event regime, this expression can be written in exponential form as
\begin{equation}
\mathrm{Prob}(X,T,N)
\simeq
1-\exp\!\left[
- N\,\frac{t_0^{\alpha}}{2\langle t\rangle}\,
T^{1-\alpha}
\int_{X/T}^{\infty}
\left(1-\frac{X}{vT}\right)
\left(\frac{z}{v}\right)^{-\alpha} f(v)\,dv
\right],
\end{equation}
where the lower bound $v>X/T$ reflects the finite-speed constraint associated with ballistic transport over a distance $X$.

Introducing the rescaled variable
\begin{equation}
z=\frac{X}{T},
\end{equation}
we define the rate function:
\begin{equation}
I(z)=\int_{z}^{\infty}
\left(1-\frac{z}{v}\right)
\left(\frac{z}{v}\right)^{-\alpha} f(v)\,dv .
\end{equation}
In terms of $I(z)$, the target-reaching probability takes the compact form
\begin{equation}
\mathrm{Prob}(X,T,N)
\simeq
1-\exp\!\left[
- N\,\frac{t_0^{\alpha}}{2\langle t\rangle}\,
T^{1-\alpha}\,I(z)
\right].
\end{equation}

This expression shows that the effect of velocity fluctuations is entirely encoded in the rate function $I(z)$. Within this large-deviation framework, redundancy acts by exponentially amplifying the probability of rare velocity-position combinations.
Importantly, the overall structure of the extreme statistics is unchanged: the probability retains the same functional dependence on $N$, $T$, and $X$, with random velocities acting only through a renormalization of the effective rate controlling rare, successful transport events. As a consequence the big-jump principle remains valid, and the extreme MFPT preserves the same scaling with the number of walkers $N$, up to non-universal prefactors determined by $f(v)$.
The extreme first-passage-time distribution is obtained by differentiation with respect to $T$,
\begin{equation}
P_N(T)
=
-\frac{\partial}{\partial T}\,
\mathrm{Prob}(X,T,N),
\end{equation}
yielding
\begin{equation}
P_N(T)
=
N\,\frac{t_0^{\alpha}}{2\langle t\rangle}\,
T^{-\alpha}
\left[
(\alpha-1)I(z)
+
z\,\frac{dI}{dz}
\right]
\exp\!\left[
- N\,\frac{t_0^{\alpha}}{2\langle t\rangle}\,
T^{1-\alpha}\,I(z)
\right].
\end{equation}

\begin{table}
\label{Table_1}
\centering
\scriptsize
\setlength{\tabcolsep}{3pt}

\begin{tabular}{@{}c c @{}}
\toprule
 \textbf{Velocity PDF $f(v)$} & \textbf{Rate function $I(z)=\int_z^\infty (1-\frac{z}{v})(\frac{v}{z})^\alpha f(v) dv$} \\
\midrule
$
f(v)=\begin{cases}
\dfrac{1}{\Delta_v}, & 0 \le v \le \Delta_v,\\
0, & \text{otherwise,}
\end{cases}
$
&
$ I(z)=\frac{\left(\frac{z}{\Delta_v}\right)^{-\alpha}
\left[-\frac{z}{\Delta_v}(\alpha+1)+\alpha\right] +\frac{z}{\Delta_v}}
{\alpha( \alpha+1)},
\qquad 0 \le z \le \Delta_v.
$\\

&

\\

$
f(v)=\frac{2}{\sqrt{2\pi}\,\Delta_v}
\exp\!\left(-\frac{v^{2}}{2\Delta_v^{2}}\right)
\ \ v\ge 0
$
&
$
I(z)=\frac{z}{\sqrt{2\pi}\Delta_v} \left[ E_{\frac{1-\alpha}{2}} \left( \frac{z^2}{2\Delta_v^2 } \right) - E_{\frac{2-\alpha}{2}} \left( \frac{z^2}{2\Delta_v^2} \right) \right] 
\qquad z\ge 0.
$
\\


&

\\

$
f(v)=\begin{cases} 
\dfrac{\beta}{\Delta_v (\beta + 1)} & 0 \le v < \Delta_v \\
\dfrac{\beta \Delta_v^\beta}{(\beta + 1) v^{1+\beta}}  & v \ge \Delta_v 
\end{cases}$
&
$I(z)=\begin{cases}
\frac{\beta \left(\frac{z}{\Delta_v}\right)}{(\alpha^2+\alpha)(\beta+1)}  + \frac{\beta \left(\frac{z}{\Delta_v}\right)^{-\alpha}}{(\alpha+1)(\beta-\alpha)}-\frac{\beta \left(\frac{z}{\Delta_v}\right)^{1-\alpha}}{\alpha(1+\beta-\alpha)}, & 0\le z < \Delta_v\\

\left(\frac{z}{\Delta_v} \right)^{-\beta}\frac{\beta}{(1+\beta-\alpha) (\beta-\alpha) (1 + \beta)}, & z\ge \Delta_v.

\end{cases}$\\

\bottomrule
\end{tabular}

\caption{Velocity distributions $f(v)$ and corresponding rate functions $I(z)$ entering the extreme first-passage statistics with random velocities, with $z=X/T$. In the first line we consider the uniform distribution with $0\leq v \leq \Delta_v$. In the second line $f(v)$ is a half Gaussian and $E_\nu(\cdot)$ denotes the exponential integral of order $\nu$. The truncated power-law case explicitly includes both the bounded contribution for $v<\Delta_v$ and the heavy-tailed contribution for $v\ge\Delta_v$. For this distribution, the only valid case has $\beta>\alpha$, for which $I(z)$ is finite and big jumps are still rare.}
\label{tab:rate_functions_landscape}
\end{table}

Although the explicit evaluation of the MFPT generally requires numerical integration, this form makes clear that velocity heterogeneity does not alter the extreme-statistics mechanism, but only reshapes the effective rate through $I(z)$.

To illustrate how different forms of motility heterogeneity map onto different rate functions, we now consider three representative choices for the velocity distribution: a uniform distribution, a Gaussian distribution, and a truncated power-law distribution, as shown in Table \ref{Table_1}. 
The case of uniform velocity distribution also describes the situation in higher dimensions in which $N$ walkers have to reach an surface boundary of a domain by projecting the velocity components on the shortest path.

\section{Extreme Value Statistics for the Position of $N$ Walkers}

We first remark that, in single-walker models where the big-jump dynamics is memoryless, the BJ approach yields identical asymptotic results for the exit time, the position, and the maximum displacement \cite{BJ_6}. Indeed, the probability of rare events for these observables is governed by the same physical mechanism, namely a single long jump of length larger than $X$ occurring within the time interval  $T-X/v$.

Therefore, when considering $N$ walkers, Eq. \eqref{probN} can be interpreted also as the equation for the probability $\text{Prob}\left(X,T,N \right)$ that the furthest walker at time $T$ lies beyond a distance $X$, i.e. the extreme value statistics of the positions of $N$ walkers. One obtains:
\begin{align}
\label{probXN}
\text{Prob}\left(X,T, N\right)&\sim
\begin{cases}
  1-\exp{\left[-\frac{N(\alpha-1)(vt_0)^{\alpha-1}}{\ 2\alpha X^{\alpha-1}}\left(\frac{vT}{X}-1 \right) \right]} & \text{if $\ell(T)\ll X\leq vT$}\\
  0 & \text{if $X\geq vT$}
\end{cases}
\end{align}
with $X\gg \ell(T)$.
Remarkably for $\ell(T)\ll X\ll VT$ the distribution exhibits a typical behavior of a Frech\'et distribution $\exp(-x^{-\alpha+1})$ with the scaling variable $x\sim X/N^{\frac{1}{\alpha-1}}$). Indeed in this regime,  the power-law tail of the single-jump distribution dominates the $N$ walker dynamics.  Conversely for $X \lesssim  vT$,  introducing the rescaled variable $X=vT- \frac{\epsilon C}{N}$ ($C=\frac{2 \alpha T^\alpha v}{(\alpha-1)t_0^{\alpha-1}}$),  the distribution crosses over to a Weibull distribution $\exp(-\epsilon)$ which is indeed, the typical behavior of systems characterized by finite support. The crossover reflects the fact that at intermediate distances $\ell(T)\ll X\ll VT$ the dynamics is ruled by the power law distribution of jump length, while at larger distances $X \lesssim  vT$ it is dominated by the cut off due to finite velocity.

Interestingly, for $\alpha<2$ the short distance ($X\ll \ell(T)$) regime is effectively diffusive, and the extreme value statistics of the positions for $N$ walkers falls into the Gumbel class. As a function of $X$ one therefore observes two crossover in the extreme value statistic: from Gumbel to Frech\'et when $X\ll \ell(T)$ and from Frech\'et to Weibull when $X$ approaches $vT$. Increasing $N$ shifts the relevant range of $X$ toward larger values, thereby revealing the different regimes of extreme-value behavior.

Let us finally return to the first-passage-time problem in the big-jump regime. As can be seen from Eq.~\eqref{probN}, the extreme-value statistics is always of the Weibull type in this case, and no Fréchet statistics is observed. Indeed, in this regime the probability of performing a big jump larger than $X$ is effectively uniform throughout the available time window, $X/v < T < L^z$.

\section{The Problem of Animal Fertilization}

We consider mammalian fertilization as a paradigmatic example of a redundancy-controlled target-search problem. In vitro sperm cells exhibit motion that can be reasonably approximated as a random walk \cite{SpermMotility_TwoStateModel_1, SpermMotility_TwoStateModel_2, SpermMotility_RandomAngleBallistic}. In vivo, however, the situation is substantially more complex and includes directed components regulated by a combination of physical and taxis mechanisms (e.g.\ uterine contractions, rheotaxis, thermotaxis and chemotaxis) \cite{miller2024sperm,eisenbach2025sperm}. 
In particular, navigation in the final segment of the reproductive tract, through the Fallopian tubes, is governed by active guidance mechanisms. Anyhow, only a very small fraction of the initial sperm population reaches this last phase.

In this setting we use a minimal first-passage description as an effective, coarse-grained tool for the statistics of long relocations generated by the interplay of motility heterogeneity and environmental transport. Within this framework, big jumps should be analized phenomenologically as rare, favorable displacement events (e.g.\ unusually long progressive runs and/or transport-assisted excursions) that dominate the early-arrival statistics. 
The purpose of the model is therefore not a microscopic description of sperm motility, but a consistency test: whether a simple heavy-tailed first-passage scenario can reproduce a robust cross-species scaling between the number $N$ of ejaculated spermatozoa and the characteristic size $X$ of the uterus across different mammalian species. 
The tail exponent $\alpha$ then encodes, in a phenomenological way, the effect of the disorder and the statistical weight of rare long relocations in the early-arrival dynamics. 

\begin{figure}
\begin{center}
\includegraphics[scale = 0.27]{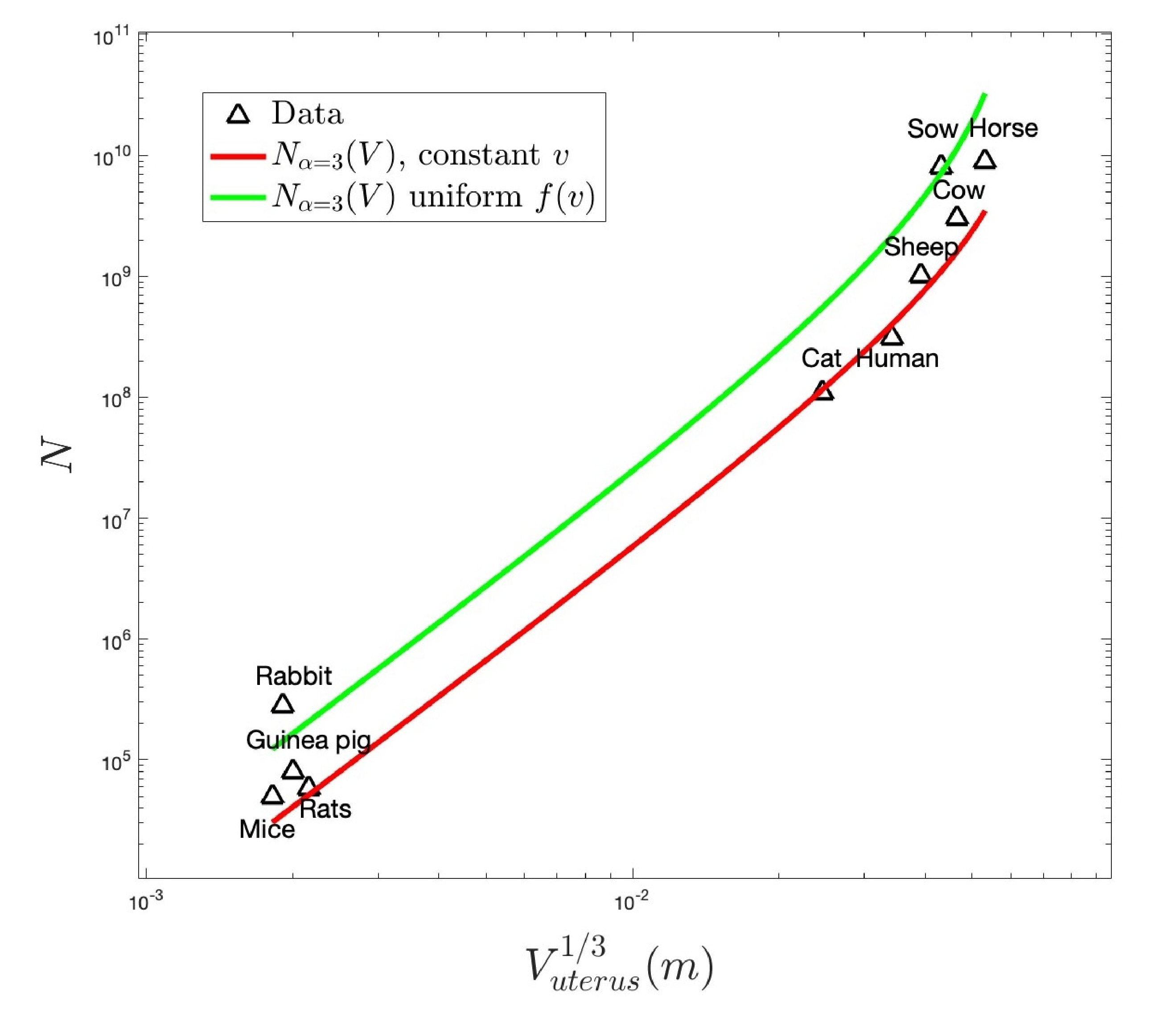}
\end{center}
\caption{Logarithmic plot of the number of swimmers as a function of the size $X \sim V^{1/3}$ of the animal uterus. 
The black triangles are experimental data taken from the review by Schuss et al.~\cite{N_Brownian_3_ReviewHolcman}, while the colored lines correspond to the $N$ HT walks model with a power exponent fixed at $\alpha=3$. 
In these simulations, the extreme MFPT is fixed to $\langle T_N \rangle = 15 \ \text{min}$, the average speed of the swimmers is $v=75 \ \mu\text{m/s}$, and the typical waiting time is $t_0=4\cdot 10^{-2} \ \text{s}$.}
\label{fig:sperm_motility}
\end{figure}

From the biological literature and in vitro experiments, the main physiological parameters can be extracted: a typical fertilization time for animal species $\langle T_N \rangle \sim 15 \ \text{min}$, an average swimming speed $v \sim 75 \ \mu\text{m/s}$, and a characteristic step duration $t_0 \sim 4 \cdot 10^{-2} \ \text{s}$, which corresponds to the typical beating time of the flagella of a sperm cell \cite{SpermMotility_1, SpermMotility_2, SpermMotility_3, SpermMotility_4}. 
Moreover, empirical data from reproductive biology report the number of swimmers $N$ as a function of uterine volume $V$ for different animal species (see Schuss et al., Fig.~10 in \cite{N_Brownian_3_ReviewHolcman}). 
Taking the uterus as a roughly isotropic three-dimensional domain, we relate its characteristic length to volume as $X \sim V^{1/3}$.

Fixing the fertilization time $\langle T_N \rangle$ and inserting the known physical and biological constants into Eq.~\eqref{eq:MFPT}, our model yields a scaling prediction for the required number of swimmers $N(V)$ across species. In Fig.~\ref{fig:sperm_motility} we show that the observed trend can be reproduced with an effective tail exponent $\alpha = 3$. 
This choice corresponds to HT walks with finite mean and variance (hence Gaussian bulk behavior by the CLT), while still allowing the extreme statistics to be dominated by the BJP mechanism. 

To improve biological realism, our model can be further generalized by allowing the speed $v$ of the walkers to vary according to a distribution $f(v)$, as discussed in the previous section. The BJP remains valid in this extended models, and its predictions are unchanged up to a renormalization of the effective speed.

Accordingly, our model suggests a possible explanation for why extreme statistics may play a central role in fertilization: in our framework, the relevant point is not that spermatozoa trajectories are microscopically L\'evy-like, but rather that rare transport events may still dominate the statistics of the earliest successful arrivals, even when the bulk dynamics appears diffusive.


\section{Conclusions}

We have developed an analytical framework to compute the extreme MFPT for $N$ independent HT walkers in finite domains. 
When rare events dominate transport, the BJP provides a natural description of target reaching, leading to a ballistic scaling $\langle T_N\rangle \sim 1/N$ that strongly outperforms classical diffusive strategies. 
This behavior highlights the central role of redundancy in shaping extreme first-passage statistics: increasing the number of walkers enhances the probability that a single rare fluctuation dominates the dynamics. Our findings are consistent with recent mathematical results for L\'evy flights ($\alpha<2$) \cite{N_Brownian_Lawley_3_Levy}, which represent idealized superdiffusive processes allowing instantaneous long jumps.
We have shown that this redundancy-driven regime is robust under generalizations such as random velocities, which renormalize the effective statistics without altering the underlying mechanism. 
At fixed $N$, we identified a crossover controlled by a critical tail exponent $\alpha_c(N)$, separating a BJP dominated regime from a Gaussian regime governed by the CLT for $\alpha>2$. 

Very recent papers \cite{New_Lawley_2021, Grebenkov_2026_New} take into account a model with constant velocity $v$ and exponentially distributed durations of the time steps. They prove the presence of a crossover from the standard Brownian behavior characterized by $1/\log(N)$ decay of the exit times to a regime where an exponential decay with $N$ to the minimum value $X/v$ is observed. This highly efficient regime emerges only under extreme redundancy, i.e., when the number of walkers grows exponentially with the system size. Our big-jump framework does not apply in this super-redundant regime. Moreover, this behavior appears to be linked to the finite probability of exiting on the first jump. It would be interesting to investigate whether a similarly rapid convergence can also arise in our setting at very large redundancy. 

Our results also suggest a connection with the inverse problem of the slowest first-passage time among many walkers \cite{Eli_Slowest_FPT}, pointing to a broader duality between extreme acceleration and extreme delays in stochastic transport.

As an application, we considered mammalian fertilization as a paradigmatic example of a redundancy-controlled search process. 
We adopted a coarse-grained first-passage perspective. 
Fixing the fertilization time leads to a scaling relation between the number of spermatozoa and the characteristic size of the uterus, which is compatible with available empirical data (Figure 10 in \cite{N_Brownian_3_ReviewHolcman}).
We expect that our results can be useful in a wide class of problems in which redundancy, combined with rare-event statistics, may provide an effective explanation to understand how fast target reaching can emerge from stochastic dynamics. Potential examples include immune cell targeting \cite{Conclusions_Cellular_1}, animal foraging strategies \cite{Conclusions_Ecology}, and intracellular transport \cite{Conclusions_Cellular_2}.

\section{Acknowledgements}

We gratefully acknowledge Vittoria Sposini for useful suggestions and stimulating discussions. R. B. is supported by the Project funded under the National Recovery and Resilience Plan (NRRP), Mission 4 Component 2 Investment 1.4—Call for tender No. 3138 of 16/12/2021 of Italian Ministry of University and Research funded by the European Union—NextGenerationEU, Award Number: Project code CN00000023, Concession Decree No. 1033 of 17/06/2022 adopted by the Italian Ministry of University and Research, CUP D93C22000400001.

\appendix

\section{Calculation for the Extreme FPT for $N$ Brownian Walkers}
\label{AppendixA}

In this section, we follow the derivation of $\langle T_N \rangle$ reported in \cite{N_Brownian_4}. Consider $N$ non interacting IID Brownian particles in a bounded domain $\Omega$, and let it be $t_i$ with $i=1,..,N$ the escape time of the $i$-th trajectory from the domain, and define $T_N$ as the shortest one, i.e.:
\begin{equation}
T_N = \min \{t_1,..., t_N\}    
\end{equation}
Now, the cumulative density function of $T_N$ for $N$ independent walkers is:
\begin{equation}
    \text{Prob}(T_N > t) = \text{Prob}(t_1>t)\cdot ... \cdot \text{Prob}(t_N>t) =\text{Prob}(t_1>t)^N
\end{equation}
On the boundary, we define absorbing regions $\partial\Omega_{a}$ and reflective regions $\partial\Omega_{r}$, and the PDF for the position of a Brownian particle $P(\vec{x},t)$ confined in the domain $\Omega$ in general dimension is:
\begin{equation}
    \frac{\partial P(\vec{x},t)}{\partial t} = D \triangle P(\vec{x},t) \ \ \ \text{for} \ \vec{x}\in \Omega, t>0,
\end{equation}
where the initial condition is $P(\vec{x},0)=P_0(\vec{x})$ for $\vec{x}\in \Omega$, $D$ is the diffusive constant, and the reflective and absorbing boundary conditions gives us other two constraints for $P(\vec{x},t)$:
\begin{equation}
    \frac{\partial P(\vec{x},t)}{\partial \vec{n}}=0 \ \ \ \text{for \ } \vec{x} \in \partial \Omega_{r}, \ \ \ \ P(\vec{x},t)=0 \ \ \ \text{for \ } \vec{x} \in \partial \Omega_{a}
\end{equation}

Then, the survival probability for a single Brownian walker is:
\begin{equation}
    \text{Prob}(t_1>t) = \int_{\Omega} P(\vec{x},t) d\vec{x}
\end{equation}

So, the first exit PDF from $\Omega$ of the fastest between $N$ Brownian particles is:
\begin{align}
P_N(T=t) &= P_N(t) = -\frac{d}{dt}\text{Prob}(T>t) = -\frac{d}{dt}\text{Prob}(t_1>t)^N \\ \nonumber
&= N (\text{Prob}(t_1>t))^{N-1}\text{Prob}(t_1=t),
\end{align}
where the probability of escaping  $\text{Prob}(t_1 =t)$ at time $t_1=t$ from the absorbing boundary, supposing it is divided in an ensemble of $N_R$ independent absorbing regions, i.e. $\partial \Omega_{a} = \cup_{i=1}^{N_R} \partial \Omega_i$, is given by the following probability flux over the boundary:
\begin{align}
    \text{Prob}(t_1>t) &= \oint_{\partial\Omega_{a}} \frac{\partial P(\vec{x},t)}{\partial \vec{n}} dS_{\vec{x}}\\ \nonumber
    &=\oint_{\partial\Omega_{1}} \frac{\partial P(\vec{x},t)}{\partial \vec{n}} dS_{\vec{x}} + ... + \oint_{\partial\Omega_{N_R}} \frac{\partial P(\vec{x},t)}{\partial \vec{n}} dS_{\vec{x}}\\ \nonumber
    &=N_R \oint_{\partial\Omega_{1}} \frac{\partial P(\vec{x},t)}{\partial \vec{n}} dS_{\vec{x}}
\end{align}

The final expression for the extreme first passage PDF of $N$ Brownian particles is:
\begin{align}
\label{probNbrown_general}
    P_N(t) &= N(\text{Prob}(t_1>t))^{N-1} \text{Prob}(t_1=t) \\ \nonumber
    &=N \left[ \int_{\Omega} P(\vec{x},t) d\vec{x} \right]^{N-1} N_R \oint_{\partial\Omega_1}\frac{\partial P(\vec{x},t)}{\partial \vec{n}} dS_{\vec{x}}
\end{align}
Let us now consider the case of a one-dimensional Brownian particle starting at zero and moving in an interval $(-\infty, X]$ with an absorbing boundary condition at $x=X$. It satisfies the usual Fokker-Planck equation, with the following constraints:
\begin{equation}
    \frac{\partial P(x,t)}{\partial t}=D \frac{\partial^2 P(x,t)}{\partial x^2} \ \ \text{for \ } x \in (-\infty,X], \ \ P(x,0)=\delta(x), \ \ P(X,t) = 0 \ \ \text{for \ } t>0,
\end{equation}

The solution of this diffusion equation is easy, and can be obtained with the usual technique of image method \cite{First_Passage_2}:
\begin{equation}
    P(x,t) = \frac{1}{\sqrt{4Dt}} \left[ \exp \left( -\frac{x^2}{4Dt} \right) - \exp \left( - \frac{(2X-x)^2}{4Dt} \right) \right]
\end{equation}

The survival probability is:
\begin{align}
    \text{Prob}(t_1>t) &= \int_{-\infty}^X P(x,t) dx \\ \nonumber
    &= 1 - \text{erfc}\left(\frac{X}{\sqrt{4Dt}}\right)\\ \nonumber
    &= 1 - \frac{2}{\sqrt{\pi}} \int_{X/\sqrt{4Dt}}^{\infty} e^{-u^2} du
\end{align}

The probability flux over the absorbing boundary condition is:
\begin{align}
-\oint_{\partial\Omega_a=\{X\}}\frac{\partial P(\vec{x},t)}{\partial \vec{n}} dS_{\vec{x}} &= -\frac{\partial P(x,t)}{\partial x} \bigg|_{x=X} \\ \nonumber
&= \frac{X}{\sqrt{4(Dt)^3}} \exp\left(-\frac{X^2}{4Dt}\right)
\end{align}

Now one can write an analytical expression for the extreme first passage time PDF of an ensemble of $N$ independent one-dimensional Brownian particles with an absorbing boundary, or target, fixed at $x=X$:
\begin{equation}
    P_N(t) = N\left[ 1 - \text{erfc}\left(\frac{X}{\sqrt{4Dt}}\right) \right]^{N-1} \frac{X}{\sqrt{4(Dt)^3}} \exp\left(-\frac{X^2}{4Dt}\right)
\end{equation}

The extreme MFPT $\langle T_N \rangle$ results from an integration by parts, and its usual relation with the survival probability $\text{Prob}(T_N>t)$:
\begin{align}
    \langle T_N \rangle &= \int_{0}^{\infty} t P_N(t) dt \\ \nonumber
    &= \int_{0}^{\infty} \text{Prob}(T_N>t) dt\\ \nonumber
    &= \int_{0}^{\infty} \text{Prob}(t_1>t)^N dt \\ \nonumber
    &= \int_{0}^{\infty} \exp\{ N \ln \text{Prob}(t_1>t) \} dt
\end{align}

Using the Laplace saddle point method, taking the limit of a large number of walkers $N \gg 1$ we can extract an analytical result for $\langle T_N \rangle$. Assuming that $X/\sqrt{4Dt}\gg 1$, i.e. at short timescales $t$ or at far target positions $X$, we obtain:
\begin{align}
    \langle T_N \rangle &= \int_{0}^{\infty} \exp \bigg\{ -N \ln \left( 1 - \frac{2}{\sqrt{\pi}} \int_{X/\sqrt{4Dt}}^{\infty} e^{-u^2} du \right) \bigg\} dt \\ \nonumber
    &\underset{N\gg1}{\sim} \int_{0}^{\infty} \exp \bigg\{ -N \frac{2}{\sqrt{\pi}} \int_{X/\sqrt{4Dt}}^{\infty} e^{-u^2} du \bigg\} dt\\ \nonumber
    &\underset{X/\sqrt{4Dt}\gg 1} {\sim} \int_{0}^{\infty} \exp \bigg\{ -N \frac{\sqrt{4Dt}}{X\sqrt{\pi}} e^{-\frac{X^2}{4Dt}} \bigg\} dt
\end{align}

If we rewrite this asymptotic estimate of the first passage time PDF $P_N(t)$ in terms of the observable $u = X/\sqrt{4Dt}$, we obtain an interesting result in the limit of $N, u \gg 1$:
\begin{equation}
    P_N(t) \underset{N,u \gg 1}{ \sim} \exp \bigg\{ -N \frac{1}{\sqrt{\pi}} \frac{e^{-u^2}}{u} \bigg\}
\end{equation}
So, the distribution of the extreme MFPT $\langle T_N \rangle$ of $N$ independent Brownian particles follows the Gumbel statistics.  After a series of change of variables and integration by parts our final result is:
\begin{equation}
    \langle T_N \rangle \approx \frac{X^2}{4D \ln \left(\frac{N}{\sqrt{\pi}} \right)} \ \ \ \ \ \text{for} \ N \gg 1, \frac{X^2}{4Dt} \gg 1
\end{equation}

\section{The Crossover Exponent $\alpha_c(N)$}\label{AppendixB}

In this section, we estimate the exponent $\alpha_c(N)$ which marks the crossover in the extreme MFPT from the optimal abundance regime to the low abundance regime in a system of $N$ walkers of size $X$. We consider the case $\alpha>2$ where the low abundance regime is described by a Brownian motion.
In particular, for $\alpha<\alpha_c(N)$, in the optimal/high abundance regimes the big jump determines the MFPT scales as $1/N$, while for $\alpha>\alpha_c(N)$ the MFPT follows a logarithmic scaling $1/\ln N$ typical of a Browninan motion.  

We will compare the extreme MFPT for $N$ independent HT walkers calculated with our BJP argument:
\begin{equation}
    \langle T_N \rangle_{HT} = \frac{X}{v} + \frac{2}{N}\frac{\alpha t_0}{(\alpha-1)}\left(\frac{X}{vt_0}\right)^{\alpha}
\end{equation}
with the extreme MFPT for $N$ independent particles following Brownian motion for $\alpha>2$:
\begin{equation}
    \langle T_N \rangle_{BM} = \frac{X^2}{4D \ln \left(\frac{N}{\sqrt{\pi}} \right)}.
\end{equation}
The diffusive constant $D$ can be calculated from  the Einstein relation:
\begin{align}
    D &= \frac{\langle X^2 \rangle}{2 \langle t \rangle} 
      = \frac{\int_{vt_0}^{\infty} X^2 p(X) dX}{2\int_{t_0}^{\infty}t p(t) dt}= \\ \nonumber
      &= \frac{\int_{vt_0}^{\infty} \alpha \frac{(vt_0)^{\alpha}}{X^{\alpha-1}} dX}{2\int_{t_0}^{\infty} \alpha \frac{(t_0)^{\alpha}}{t^{\alpha}} dt} = \frac{\alpha-1}{\alpha-2}\frac{v^2 t_0}{2}
\end{align}
where we assume that the step duration $t$ follows the power law distribution $p(t) \sim \alpha t_0^{\alpha} t^{-1-\alpha} \theta(t-t_0)$, and the finite velocity is constant  i.e. $X=vt$.
Now, to find the crossover value of $\alpha$ for the transition between the two regimes to occur, we set the equation $\langle T_N \rangle_{BM} = \langle T_N \rangle_{LW}$ and solve it in terms of $\alpha$. We obtain an implicit equation of $\alpha$ as a function of $N$. In our derivation, we assume that the target size $X$, the speed $v$, and the typical waiting time $t_0$ are fixed parameters. Therefore:
\begin{align}
     \langle T_N \rangle_{BM} &= \langle T_N \rangle_{HT}\\ \nonumber 
    \frac{X^2}{4D \ln \left(\frac{ N}{\sqrt{\pi}} \right)} &= \frac{X}{v} + \frac{2}{N}\frac{\alpha t_0}{(\alpha-1)}\left(\frac{X}{vt_0}\right)^{\alpha} \\ \nonumber
    \frac{(\alpha-2)(X/v)^2}{2 (\alpha-1)t_0\ln \left(\frac{ N}{\sqrt{\pi}} \right)} &= \frac{X}{v} + \frac{2}{N}\frac{\alpha t_0}{(\alpha-1)}\left(\frac{X}{vt_0}\right)^{\alpha} \\ \nonumber
    \frac{N}{2\alpha}\left(\frac{(\alpha-2)(X/v)}{2t_0 \ln\left(\frac{ N}{\sqrt{\pi}} \right)}-1\right)&= \left(\frac{X}{vt_0}\right)^{\alpha-1} \\ \nonumber
    \ln \left[ \frac{N}{2\alpha}\left(\frac{(\alpha-2)(X/v)}{2t_0 \ln\left(\frac{N}{\sqrt{\pi}} \right)}-1\right)\right] &= (\alpha-1) \ln \left(\frac{X}{vt_0}\right)\\ \nonumber
    \ln \left[ \frac{N}{2\alpha}\left(\frac{X}{vt_0}\right)\left(\frac{(\alpha-2)}{2\ln\left(\frac{ N}{\sqrt{\pi}} \right)}\left(\frac{X}{vt_0}\right)-1\right)\right] &= \alpha \ln \left(\frac{X}{vt_0}\right)
\end{align}

Let us define the auxiliary variable $u=\alpha \ln r$, where $r=X/vt_0$, and rewrite the equation in terms of $u$:
\begin{align}
    \ln \left[ \frac{N}{2\alpha}r\left(\frac{(\alpha-2)}{2\ln\left(\frac{ N}{\sqrt{\pi}} \right)}r-1\right)\right] &= u \\ \nonumber
    \frac{N}{2\alpha}r\left(\frac{(\alpha-2)}{2\ln\left(\frac{ N}{\sqrt{\pi}} \right)}r-1\right) &= e^u \\ \nonumber
    \frac{N}{2} r \ln r \left(\frac{(\alpha-2)}{2\ln\left(\frac{ N}{\sqrt{\pi}} \right)}r-1\right) &= u e^u
\end{align}
where in the last line we recognize the transcendental equation $z=ue^u$ that defines the principal branch of the Lambert function $u=W_0(z)$, since $z>0$. Therefore, remembering that $u=\alpha \ln r$, dividing both sides of the equation by $\ln r$ we finally obtain the implicit relation for $\alpha_c$, which is only a function of $N$ and $\alpha_c$ itself. This equation can be solved numerically and does not have an analytical solution, and coincides with the one reported in the paper:

\begin{align}
\label{alpha_crit_exact}
    \alpha_c &= \frac{1}{\ln \left( \frac{X}{vt_0} \right)} W_0 \left(\frac{N}{2}  \left( \frac{X}{vt_0} \right)\ln \left( \frac{X}{vt_0} \right) \left[\frac{(\alpha_c-2)}{2\ln\left(\frac{ N}{\sqrt{\pi}} \right)}\left( \frac{X}{vt_0} \right)-1\right]\right) \\
    &\equiv \frac{W_0(\mathcal{C}_{X,v,t_0}(N, \alpha_c))}{\ln \left( \frac{X}{vt_0} \right)}
\end{align}

We conclude this section by showing the asymptotic behavior of $\alpha_c(N)$ for $N \gg 1$. In this asymptotic limit, we compare the logarithmic scaling in $N$ of $\langle T_N \rangle_{BM}$ with the power scaling in $N$ of $\langle T_N \rangle_{HT}$, in particular neglecting the constant ballistic term $X/v$, i.e. $\langle T_N \rangle_{HT} \sim \frac{2}{N}\frac{\alpha t_0}{\alpha-1}\left( \frac{X}{vt_0} \right)^{\alpha}$. Comparing the two extreme MFPTs, we observe that:

\begin{align}
 \frac{X^2}{4D \ln \left(\frac{ N}{\sqrt{\pi}} \right)}& \sim \frac{2}{N}\frac{\alpha t_0}{\alpha-1}\left( \frac{X}{vt_0} \right)^{\alpha} \\ \nonumber
 \frac{N}{\ln \left(\frac{ N}{\sqrt{\pi}} \right)} \frac{\alpha-2}{4\alpha} &\sim \left(\frac{X}{vt_0}\right)^{\alpha-2} \\ \nonumber
 \ln \left[ \frac{N}{\ln \left(\frac{ N}{\sqrt{\pi}} \right)} \frac{\alpha-2}{4\alpha}\right] &\sim (\alpha-2) \ln \left(\frac{X}{vt_0}\right) \\ \nonumber
 \ln N -\ln \left(\ln \left(\frac{N}{\sqrt{\pi}}\right) \right) + \ln\left(\frac{\alpha-2}{4\alpha}\right) &\sim (\alpha-2) \ln \left(\frac{X}{vt_0}\right)
\end{align}

\begin{figure}
\begin{center}
\includegraphics[scale = 0.7]{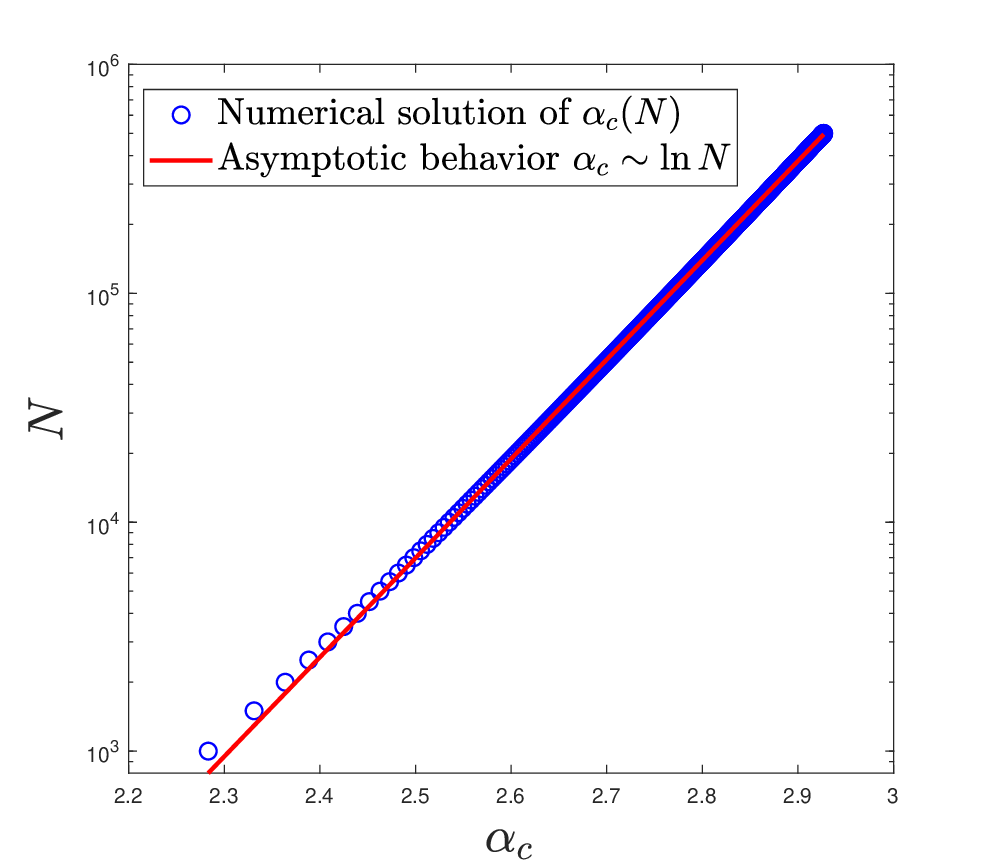}
\end{center}
\caption{Logarithmic plot of the crossover tail index $\alpha_c$ as a function of the number of walkers $N$. The blue circles are data calculated by solving numerically the implicit relation for $\alpha_c$ (Eq. \eqref{alpha_crit_exact}). The red line is the asymptotic behavior of $\alpha_c$ in the limit of large $N$ (Eq. \eqref{alpha_crit_approx}). In this simulation, the ratio between the target position and the typical length scale is fixed to $X/vt_0 = 25000$;}
\label{fig:alpha_approx}
\end{figure}

Since we are in the limit of $N \gg 1$, we can neglect the subleading term scaling log-logarithmically in $N$, i.e. $\ln (\ln N)$ and the constant logarithmic term in alpha, i.e. $\ln((\alpha-2)/4\alpha)$. So, the asymptotic behavior of $\alpha_c$ is:
\begin{equation}
\label{alpha_crit_approx}
    \alpha_c(N) \underset{N \gg 1}{\sim} 2 + \frac{\ln N}{\ln \left( \frac{X}{vt_0}\right)}
\end{equation}

This logarithmic behavior in the large $N$ limit can be easily checked by numerical simulations, see Figure \ref{fig:alpha_approx}.

\section{The BJP and the extreme MFPT for $\alpha \rightarrow 1^+$}\label{AppendixC}

In this section, we show that the discrepancy observed in Fig. 2 of the main work for the MFPT when $\alpha \rightarrow 1^+$ between the numerical experiments and our calculation, is due the fact that in this regime for too small sizes we cannot apply the BJP even for a single walker.
In particular the BJP holds if the system size $X$ is much larger than
the typical distance $\ell(T)$ covered by the process with many jumps in a time $T=X/v$ (i.e. $T$ is the duration of the single big jump which exit from the system at constant velocity $v$).

The characteristic length $\ell(T)$ is defined by the scaling form of the PDF $P(x,t)$ of a single walker. We start from its Fourier-Laplace transform $\tilde{P}(k,s)$, following the standard approach shown in \cite{LevyWalks_1, AppendixC_ref0, AppendixC_ref1}, 
which at constant velocity is given by:

\begin{equation}
\label{montrollweiss}
    \tilde{P}(k,s) = \frac{\frac{1}{2}\left[\frac{1-\tilde{\psi}(s+ivk)}{s+ikv}+ \frac{1-\tilde{\psi}(s-ivk)}{s-ikv} \right]}{1-\frac{1}{2}(\tilde{\psi}(s+ivk)+\tilde{\psi}(s-ivk))}
\end{equation}
where $\tilde{\psi}(s)$ is the Laplace transform of the power-tailed PDF for waiting times $\psi(t) = \alpha t_0^{\alpha}/t^{1+\alpha}$, with $\psi(t<t_0)=0$.

To obtain the scaling law of $P(x,t)$, we perform an asymptotic estimate of its Fourier-Laplace transform in the limit of $k,s \rightarrow 0$. To do this, we start again from the various terms of the Montroll-Weiss formula, and begin by developing $\tilde{\psi}(s)$ for $s \rightarrow 0$, assuming $1<\alpha<2$:
\begin{equation}
    \tilde{\psi}(s+ikv) \underset{s,k\rightarrow0}{\sim} 1 - (s+ikv) \langle t \rangle + B_{\alpha} (s+ikv)^{\alpha} + o((s+ikv)^{\alpha}) 
\end{equation}
with $B_{\alpha}=\Gamma(1-\alpha)t_0^{\alpha}<0$.
We can expand the square bracket in the numerator up to the first order in $s$ and $k$ obtaining that the numerator in the limit of $k,s\to 0$ tends to the constant $\langle t \rangle$. Similarly, we can obtain the expression for $k,s \rightarrow 0$ of the denominator:
\begin{align}
     \frac{1}{2} [\tilde{\psi}(s-ikv) + \tilde{\psi}(s+ikv)] 
    &\underset{k,s\rightarrow 0}{\sim} 1-s\langle t\rangle -\frac{B_{\alpha}}{2}[(s-ikv)^{\alpha} + (s+ikv)^{\alpha}]  \\ \nonumber
    &\underset{k,s\rightarrow 0}{\sim} 1-s + \frac{B_{\alpha}}{2}v^{\alpha}k^{\alpha}[(-i)^{\alpha} + (+i)^{\alpha}] 
\end{align}
Where within the limit of $k, s \rightarrow 0$, $s^\alpha$ can be neglected with respect $s$, so we can approximate $(s \pm ikv)^{\alpha} \sim  k^{\alpha}(\pm i)$. Our formula becomes:
\begin{align}
    \tilde{P}(k,s) &\underset{k,s \rightarrow 0}{\sim}\frac{1}{s + \frac{B_{\alpha}}{2\langle t \rangle}v^{\alpha}k^{\alpha}[(-i)^{\alpha} + (+i)^{\alpha}]}\\ \nonumber
    &\underset{k,s \rightarrow 0}{\sim} \frac{1}{s - \frac{B_{\alpha}}{2\langle t \rangle}v^{\alpha}k^{\alpha} \cos\left(\frac{\alpha \pi}{2}\right)}\\ \nonumber
    &\underset{k,s \rightarrow 0}{\sim} \frac{1}{s + D_{\alpha}k^{\alpha}}
\end{align}

where the coefficient $D_{\alpha}$ is defined as:

\begin{equation}
    D_{\alpha} =  -\frac{B_{\alpha}v^{\alpha}}{\langle t \rangle} \cos \left( \frac{\alpha \pi}{2} \right) 
    =-v^{\alpha}t_0^{\alpha-1}\Gamma(1-\alpha)\frac{\alpha-1}{\alpha}>0
\end{equation}
}
At this point, we recognize that the Fourier-Laplace transform $\tilde{P}(k,s)$ has a scaling law of the form:

\begin{equation}
    \tilde{P}(k,s) = \frac{1}{s} g\left( \frac{k}{s^{1/\alpha}} \right) \ \ \ \text{with \ } g\left(\frac{k}{s^{1/\alpha}}\right) =\frac{1}{1 + D_{\alpha}\left(\frac{k}{s^{1/\alpha}}\right)^{\alpha}}
\end{equation}

Now, to arrive at the asymptotic expression of the PDF in real space $P(x,t)$, we perform the inverse Fourier-Laplace transform on our dual PDF $\tilde{P}(k,s)$. We start from the inverse Laplace transform on $s$, which returns:
\begin{align}
    \hat{P}(k,t) &= \mathcal{L}^{-1}(\tilde{P}(k,s))(t)\\ \nonumber
    &= \int_0^{\infty} \frac{e^{st}}{s+D_{\alpha}k^{\alpha}}ds \\ \nonumber
    &= e^{-D_{\alpha}|k|^{\alpha}t}
\end{align}
We can easily recognize that the dual PDF $\hat{P}(k,t)$ is the Fourier transform of an $\alpha$-stable L\'evy distribution, where the modulus $|.|$ on $k$ highlights the symmetry around zero of our HT walker. Therefore, we can obtain the final scaling form of the PDF $P(x,t)$ by performing the inverse Fourier transform:
\begin{align}
    P(x,t) &= \mathcal{F}^{-1}(\hat{P}(k,t))(x)\\ \nonumber
    &=\frac{1}{2\pi} \int_{-\infty}^{+\infty} e^{-ikx}e^{-D_{\alpha}|k|^{\alpha}t} dk\\ \nonumber
&\underset{z^{\alpha}=D_{\alpha}|k|^{\alpha}t}{=} \frac{1}{(D_{\alpha }t)^{1/\alpha}}\frac{1}{2\pi} \int_{-\infty}^{+\infty} e^{-iz\frac{x}{(D_{\alpha}t)^{1/\alpha}}} e^{-|z|^{\alpha}} dz
\end{align}
where, once the variable change has been made, we recognize that the integral multiplied by the factor $1/2\pi$ coincides with the formal definition of an $\alpha$-stable L\'evy distributed $L_{\alpha}(.)$, thus arriving at the final scaling form:
\begin{equation}
    P(x,t) = \frac{1}{C_{\alpha}t^{1/\alpha}} L_{\alpha}\left(\frac{x}{C_{\alpha}t^{1/\alpha}}\right) \ \ \ \text{with \ } C_{\alpha}=vt_0^{\frac{\alpha-1}{\alpha}}\left[ \Gamma(1-\alpha)\frac{\alpha-1}{\alpha} \cos\left(\frac{\alpha\pi}{2}\right)\right]^{1/\alpha}
\end{equation}

So we have proven that the typical length of HT walkers in the bulk regime scales as $\ell(t) \sim C_{\alpha}t^{1/\alpha}$. Now, if we want to stay in the BJP regime in the limit where $\alpha \rightarrow 1^+$, we need to impose that rare events driven by ballistic jumps scale much faster than the typical length $\ell(t)$. If our target is at position $X$ in a total measurement time $T$, for $\alpha \rightarrow 1^+$ it must be true that $X \gg \ell(T=X/v)$, so:
\begin{align}
    X &\gg C_{\alpha}\left(\frac{X}{v}\right)^{1/\alpha} \\ \nonumber
    X &\gg \left[ \Gamma(1-\alpha)\frac{\alpha-1}{\alpha} \cos\left(\frac{\alpha\pi}{2}\right)\right]^{1/\alpha}vt_0^{\frac{\alpha-1}{\alpha}}\left(\frac{X}{v}\right)^{1/\alpha} \\ \nonumber
    \frac{1}{\left[ \Gamma(1-\alpha)\frac{\alpha-1}{\alpha} \cos\left(\frac{\alpha\pi}{2}\right)\right]^{1/\alpha}} &\gg \left(\frac{X}{vt_0}\right)^{1/\alpha-1}
\end{align}
Now, taking the limit for $\alpha \rightarrow 1^+$, it can be easily verified that the term in square brackets tends to one, i.e. one can asymptotically verify that $C_{\alpha} \sim v t_0^{1-1/\alpha}$. So, we finally obtain a final relationship for $\alpha$:
\begin{align}
    1 &\underset{\alpha \rightarrow 1^+}{\gg} \left(\frac{X}{vt_0}\right)^{1/\alpha-1} \\ \nonumber
    1 &\underset{\alpha \rightarrow 1^+}{\gg} e^{{\left(\frac{1}{\alpha} -1\right) \ln \left(\frac{X}{vt_0}\right)}}\\ \nonumber
\end{align}

Now, introducing an arbitrary positive parameter $C > 1$ that guarantees asymptotically that the argument of the exponent is certainly much bigger than one, we can write the following inequality:

\begin{equation}
        C \underset{\alpha \rightarrow 1^+}{<}  \left(1- \frac{1}{\alpha}\right) \ln \left(\frac{X}{vt_0}\right)
\end{equation}

Therefore, for $\alpha \rightarrow 1^+$, the necessary condition to satisfy the BJP, and consequently for applying our result for the extreme MFPT $\langle T_N \rangle$, is that the following asymptotic inequality holds for $\alpha$ with respect to the threshold $X$:
\begin{equation}
    \alpha(X) \underset{\alpha\rightarrow 1^+}{>} \frac{\ln \left(\frac{X}{vt_0}\right)}{\ln \left(\frac{X}{vt_0}\right) - C}
\end{equation}

We can see that, the more we increase the threshold value $X$, the closer we can get to the asymptotic limit $\alpha \rightarrow 1^+$, and so we recover Eq. \eqref{alpha_near_1}, and this fact is confirmed by numerical simulations in Figure \ref{fig:alpha_near_1}.

\bibliographystyle{elsarticle-harv}
\bibliography{BJ_N_walkers}

\end{document}